\documentclass[pre,superscriptaddress,twocolumn,showpacs]{revtex4}
\usepackage{graphicx,bbm}
\usepackage{hyperref}
\usepackage{amsmath}
\def\etal#1{ {\em et al.}}
\def\tit#1{}
\def\ii{{\rm i}}
\def\lan{\langle}
\def\ran{\rangle}
\def\llan{\left\langle}
\def\rran{\right\rangle}
\makeatletter
\def\underbracket{\@ifnextchar [ {\@underbracket} {\@underbracket [\@bracketheight]}}
\def\@underbracket[#1]{\@ifnextchar [ {\@under@bracket[#1]} {\@under@bracket[#1][0.4em]}}
\def\@under@bracket[#1][#2]#3{
           \mathop {\vtop {\m@th \ialign {##\crcr $\hfil \displaystyle {#3}\hfil $%
                              \crcr \noalign {\kern 3\p@ \nointerlineskip }\upbracketfill {#1}{#2}
                              \crcr \noalign {\kern 3\p@ }}}}\limits}
\def\upbracketfill#1#2{$\m@th \setbox \z@ \hbox {$\braceld$}
                  \edef\@bracketheight{\the\ht\z@}\bracketend{#1}{#2}
                  \leaders \vrule \@height #1 \@depth \z@ \hfill
                  \leaders \vrule \@height #1 \@depth \z@ \hfill \bracketend{#1}{#2}$}
\def\bracketend#1#2{\vrule height #2 width #1\relax}
\def\downbracketfill#1#2{$\m@th \setbox \z@ \hbox {$\braceld$}
                  \edef\@bracketheight{\the\ht\z@}\downbracketend{#1}{#2}
                  \leaders \vrule \@height #1 \@depth \z@ \hfill
                  \leaders \vrule \@height #1 \@depth \z@ \hfill
\downbracketend{#1}{#2}$}
\def\downbracketend#1#2{\vrule depth #2 width #1\relax}
\makeatother

\begin{document}
\pagenumbering{arabic}

\title{Spectral density of a Wishart model for nonsymmetric Correlation Matrices} 
\author{Vinayak}
\email{vinayaksps2003@gmail.com}
\affiliation{Instituto de Ciencias F\' isicas, Universidad Nacional Aut\' onoma de M\' exico, C.P. 62210 Cuernavaca, M\' exico}
\date{\today}

\begin{abstract}
The Wishart model for real symmetric correlation matrices is defined as $\mathsf{W}=\mathsf{AA}^{t}$, where matrix $\mathsf{A}$ is usually a rectangular Gaussian random matrix and $\mathsf{A}^{t}$ is the transpose of $\mathsf{A}$. Analogously, for nonsymmetric correlation matrices, a model may be defined for two statistically equivalent but different matrices $\mathsf{A}$ and $\mathsf{B}$ as $\mathsf{AB}^{t}$. The corresponding Wishart model, thus, is defined as $\mathbf{C}=\mathsf{AB}^{t}\mathsf{BA}^{t}$. We study the spectral density of $\mathbf{C}$ for the case when $\mathsf{A}$ and $\mathsf{B}$ are not statistically independent. The ensemble average of such nonsymmetric matrices, therefore, does not simply vanishes to a null matrix. In this paper we derive a Pastur self-consistent equation which describes spectral density of large $\mathbf{C}$. We complement our analytic results with numerics.
\end{abstract}

\maketitle
\renewcommand*\thesection{\Roman{section}}
\renewcommand*\thesubsection{\thesection.\Roman{subsection}}

\section{Introduction}

Correlation matrices are a fundamental tool for multivariate time series analysis \cite{Wilks, Muirhead}. Wishart introduced random matrices of $\mathsf{W}=\mathsf{AA}^{t}/T$ type as a model for real symmetric correlation matrices \cite{Wishart}. In this model $\mathsf{A}$ is usually a rectangular matrix of dimension $N\times T$, $\mathsf{A}^{t}$ is the transpose of $\mathsf{A}$ and the matrix entries $A_{j\nu}$ are independent Gaussian variables with zero mean and a fixed variance. In due course this model gained much attention from various branches of science, and now is applied to a vast domain including mathematical statistics \cite{Wilks, Muirhead}, physics \cite{Mehta,Brody81,GuhrGW98, BenRMP97}, communication engineering \cite{Muller:Review}, econophysics\cite{Finance1, Finance2, Finance3, Finance4, Thomas2012, vrt2013}, biological sciences \cite{gene,Seba:03}, atmospheric science \cite{diverseAT}, etc. In random matrix theory (RMT) \cite{Mehta} ensembles of such nonnegative matrices are known as Wishart orthogonal ensembles (WOE) or Laguerre orthogonal ensembles \cite{Laguerre,APSG} where analytical results for spectral statistics are known in great detail. 

WOE characterizes a {\it null hypothesis} for symmetric correlation matrices where the spectral statistics set a benchmark or a reference against which any useful {\it actual} correlation must be viewed. For instance, the spectral density of WOE \cite{marchenko}, has proved to be remarkably useful for identifying underlying actual correlations. Prime examples thereof are found in econophysics \cite{Finance1, Finance2, Finance3}. However, there have been important advances incorporating actual correlations in random matrix ensembles. These ensembles are often referred to as the correlated Wishart ensembles \cite{SenM,Burda:2005,vp2010,guhr1, guhr2}; the generalization for the WOE case is the correlated Wishart orthogonal ensemble (CWOE). CWOE results are also important because these supply a reference against which the correlations lying off the trend must be viewed \cite{Potter:2005}

In recent years there has been a growing interest in the analysis of nonsymmetric correlation matrices \cite{John,Bouchaud:2009}. A nonsymmetric correlation matrix can be realized for time-lagged correlations among the variables representing the same statistical system \cite{Bouchaud:2007,Stanley:2011}. In a more general case it can be the matrix representing correlations among the variables of two different statistical systems \cite{Livan:2012}. The corresponding random matrix model which describes a null hypothesis for nonsymmetric correlation matrices is defined for two statistically equivalent but independent matrices, e.g., $\Sigma^{\text{AB}}=\mathsf{AB}^{t}/T$ where $\mathsf{A}$ and $\mathsf{B}$ are rectangular matrices of dimensions $N\times T$ and $M\times T$ respectively, and entries of both the matrices are the Gaussian variables with mean zero and variance one. For rectangular $\Sigma^{\text{AB}}$, i.e., for $N\ne M$, a reference against which the actual correlations have to be viewed is the statistics of singular values of $\Sigma^{\text{AB}}$ \cite{Muller,Bouchaud:2007,Burda:2010} or equivalently the statistics of eigenvalues of a corresponding Wishart model of $N\times N$ matrices, defined as $\mathbf{C}=\mathsf{AB}^{t}\mathsf{BA}^{t}/T^{2}$. On the other hand, for square $\Sigma^{\text{AB}}$, ample amount of results are known for the statistics of complex eigenvalues \cite{Burda:2010,nonhermitian1, nonhermitian2} which could be useful for the spectral studies of correlation matrices such as the eigenvalue density used in \cite{Livan:2012}.

Following the CWOE approach we generalize the Wishart model for nonsymmetric correlation matrices to the case where $\mathsf{A}$ and $\mathsf{B}$ are not statistically independent. In other words, the ensemble average $\overline{\mathsf{AB}^{t}}/T=\eta$ where bar denotes the ensemble averaging and $\eta$ is an $N\times M$ correlation matrix which defines the nonrandom correlations between $N$ input variables with $M$ output variables. The joint probability density of the matrix elements $\mathsf{A}$ and $\mathsf{B}$ can be described as
\begin{equation}\label{JPDAB}
P(\mathsf{A},\mathsf{B})\propto \exp
\left[-\text{tr}\,\left\{
\left(
\begin{matrix}
\mathbf{1} & \eta
\\
\eta^{t} & \mathbf{1}
\end{matrix}
\right)
^{-1}
\left(
\begin{matrix}
\mathsf{A} \\
      \mathsf{B} 
\end{matrix}
\right)
\left(
\begin{matrix}
\mathsf{A}^{t} \, \mathsf{B}^{t} 
\end{matrix}
\right)
\right\}
\right],
\end{equation}
where $\mathbf{1}$ is an identity matrix of dimensions $N\times N$ in the upper diagonal block and $M\times M$ in the lower diagonal block. In this paper we derive a self-consistent Pastur equation which describes the spectral density for large $\mathbf{C}$ where $\eta\ne 0$.
 
In the next section we define the nonsymmetric correlation matrices from CWOE approach, fix notations and describe some generalities of the work. In section \ref{secResult}, we state the main result of the paper. Details of the derivation of the result is given in Appendix \ref{apB}. In Sec. \ref{Numerics} we present some numerical examples to complement our result. This is followed by conclusion.

\section{Nonsymmetric Correlation Matrices: A CWOE perspective}\label{secCWE}
CWOE is an ensemble of real symmetric matrices of type $\mathcal{C}=\mathcal{WW}^{t}/T$ where $\mathcal{W}=\xi^{1/2}\mathsf{W}$, $\xi$ is a positive definite nonrandom matrix and entries of the matrix $\mathsf{W}$ are independent Gaussian variables with mean zero and variance once. Thus 
\begin{equation}
\overline{\mathcal{C}}=\xi.
\end{equation}
Spectral statistics of CWOE have been addressed by several authors. For example, the spectral density for large matrices has been derived by \cite{SenM, Burda:2005, vp2010, marchenko, Silverstien} and recently for finite dimensional matrices by \cite{guhr1, guhr2}. Unlike WOE spectra, CWOE spectra may exhibit nonuniversal spectral statistics as noted in \cite{vp2010}.

Suppose the matrix $\mathcal{W}$ constitutes of two different random matrices $\mathcal{A}$ and $\mathcal{B}$, as 
\begin{eqnarray}
\mathcal{W}=
\left(
\begin{matrix}
\mathcal{A} \\
\mathcal{B} 
\end{matrix}
\right),
\end{eqnarray}
where $\mathcal{A}$ and $\mathcal{B}$ are of dimensions $N\times T$ and $M\times T$, respectively. Then the matrix $\mathcal{C}$ is a partitioned matrix, defined in terms of $\mathcal{A}$ and $\mathcal{B}$, as
\begin{equation}\label{CWEC}
\mathcal{C}=\dfrac{1}{T} 
\left(
\begin{matrix}
\mathcal{AA}^{t} & \mathcal{AB}^{t}\\
\mathcal{BA}^{t} & \mathcal{BB}^{t}.
\end{matrix}
\right),
\end{equation}
Here the diagonal blocks viz., $\mathcal{AA}^{t}$ and $\mathcal{BB}^{t}$, and the off-diagonal blocks viz., $\mathcal{AB}^{t}$ and $\mathcal{BA}^{t}$, are respectively $N\times N$, $M\times M$, $N\times M$ and $M\times N$ dimensional. Therefore $\xi$ is also partitioned:
\begin{equation}\label{CWEXI}
\xi=
\left( 
\begin{matrix}
\xi_{\text{AA}} & \xi_{\text{AB}}\\
\xi_{\text{BA}} & \xi_{\text{BB}}
\end{matrix}
\right),
\end{equation}
where diagonal blocks $\xi_{\text{AA}}$ and $\xi_{\text{BB}}$ account for the correlations among the variables of $\mathcal{A}$ and of $\mathcal{B}$, respectively. Off-diagonal blocks, e.g., $\overline{\mathcal{AB}^{t}}/T=\xi_{\text{AB}}$, account for the correlations of $\mathcal{A}$ and $\mathcal{B}$. By construction $\xi_{\text{BA}}=[\xi_{\text{AB}}]^{t}$. Without loss of generality we consider $M\ge N$ and $T\ge M$ .

We consider the case where $\xi_{\text{AB}}\ne 0$ and wish to compare the spectral density with that of the null hypothesis, i.e., when $\xi_{\text{AA}}=\mathbf{1}_{N\times N}$, $\xi_{\text{BB}}=\mathbf{1}_{M\times M}$ and $\xi_{\text{AB}}=0$. It is therefore important to remove the cross-correlations among the variables of individual matrices because only then the diagonal blocks of (\ref{CWEC}) will yield identity matrices on the ensemble averaging. Thus we introduce decorrelated matrices \cite{vp2010} defined as 
\begin{eqnarray}\label{decorr}
\mathsf{A}&=&\xi_{\text{AA}}^{-1/2}\mathcal{A},
\nonumber\\
\mathsf{B}&=&\xi_{\text{BB}}^{-1/2}\mathcal{B}.
\end{eqnarray}
Note that we still have $\overline{\mathsf{AB}^{t}}/T=\eta$ where $\eta=\xi_{\text{AA}}^{-1/2}\,\xi_{\text{AB}}\,\xi_{\text{BB}}^{-1/2}$ and the null hypothesis is characterized for $\eta=0$. This case has been studied by several authors \cite{Muller,Bouchaud:2007, Burda:2010}. In this paper we consider $\eta\ne0$ and calculate ensemble averaged spectral density of $N\times N$ symmetric matrices $\mathbf{C}$, defined as
\begin{equation}\label{C}
\mathbf{C}=\dfrac{\mathsf{AB}^{t}\mathsf{BA}^{t}}{T^{2}}.
\end{equation}
We further define $M\times M$ symmetric matrix $\mathbf{D}$,
\begin{equation}\label{D}
\mathbf{D}=\dfrac{\mathsf{BA}^{t}\mathsf{AB}^{t}}{T^{2}},
\end{equation}
and the ratios, 
\begin{eqnarray}
\kappa_{N}&=&\dfrac{N}{T},
\\
\kappa_{M}&=&\dfrac{M}{T}.
\end{eqnarray}

A few remarks are immediate. At first we note that the ensemble averages yield, $\overline{\mathbf{C}}=\kappa_{M}\textbf{1}_{N\times N}+\eta\eta^{t}$ and $\overline{\mathbf{D}}=\kappa_{N}\,\textbf{1}_{M\times M}+\eta^{t}\eta$. By construction it is obvious that the nonzero eigenvalues of $\mathbf{D}$ are identical to those of $\mathbf{C}$. Next, for $T\rightarrow\infty$, since $\mathcal{C}=\xi$, $\mathbf{C}=\eta\eta^{t}$ and $\mathbf{D}=\eta^{t}\eta$.  We finally define a symmetric matrix $\zeta$, as
\begin{equation}
\zeta=\eta\eta^{t}.
\end{equation}

However, in the following we consider a large $N$ limit where $N/T$ and $M/T$ are finite so that matrices $\mathbf{C}$ and $\mathbf{D}$ will never be deterministic. We consider only those cases where the spectrum of $\zeta$ does not exceed $N$. This is always valid for our model because the positive definiteness of $\xi$ ensures an upper bound $1$ for eigenvalues of $\zeta$. For the completeness of the paper we prove this remark in Appendix \ref{apA}.


\section{Spectral Density for large matrices}\label{secResult}

To obtain the spectral density, $\rho(\lambda)$, of $\mathbf{C}$ we closely follow the binary correlation method developed by French and his collaborators \cite{ap81,Brody81} and used in \cite{vp2010} to study the spectral statistics of CWOE. In this method we deal with the resolvent or the Stieltjes transform of the density, defined for a complex variable $z$ as 
\begin{equation}
{G}(z)=\llan{ \dfrac{1}{z  \textbf{1}_{N\times N} -\mathbf{C}}\,}\rran_{N}.
\end{equation}
We use the angular brackets to represent the spectral averaging, e.g., $\llan H\rran_{k}= k^{-1}\,\text{tr}\, H$, and bar to denote averaging over the ensemble. Below we use bar also to represent functions of ensemble averaged scalar quantities. 

The ensemble averaged spectral density, can be determined via
\begin{equation}\label{resden}
\overline{\rho}(\lambda)=\mp \dfrac{1}{\pi}\, \Im \overline{G}(\lambda\pm \ii \epsilon), 
\end{equation}
for infinitesimal $\epsilon>0$. For large $z$, the resolvent may be expressed in terms of moments, $\overline{\mathbf{m}}_{p}$, of the density, as
\begin{eqnarray}\label{ResMp}
\overline{G}(z)&=&\sum_{p=0}^{\infty}\dfrac{\llan\overline{ \mathbf{C}^{p}}\rran_{N}}{z^{p+1}}
\nonumber\\
&=& \sum_{p=0}^{\infty}\dfrac{\overline{\mathbf{m}_{p}}}{z^{p+1}}.
\end{eqnarray}
The remaining task now is to obtain a closed form of this summation. To obtain a closed form, in the above expansion we consider only those binary associations yielding leading order terms and avoid those resulting in terms of $\mathcal{O}(N^{-1})$. This method finally gives the so called Pastur self-consistent equation for the resolvent \cite{Brody81,vp2010,ap81}, or the Pastur density \cite{Pastur}, which holds for large matrices.
 
In order to do the ensemble averaging we use the following exact results, valid for arbitrary fixed matrices $\Phi$ and $\Psi$:
\begin{eqnarray}\label{Iden1}
\dfrac{1}{T} 
\lan\overline{ \mathsf{A} \Phi \mathsf{A}^{t}\Psi}
\ran_{N}
&=&\llan \Phi \rran_{T} \llan \Psi \rran_{N},
\\
\label{Iden2}
\lan\overline{ \mathsf{A} \Phi \mathsf{A}\Psi}
\ran_{N}
&=&\llan \Phi^{t}\Psi\rran_{N},
\\
\label{Iden3}
\overline{\llan \mathsf{A} \Phi\rran_{N} \llan \Psi \mathsf{A}^{t}
\rran_{N}}
&=&\dfrac{1}{N}
\llan \Psi \Phi \rran_{N},
\\
\label{Iden4}
\overline{\llan \mathsf{A} \Phi\rran_{N} \llan \mathsf{A} \Psi
\rran_{N}}
&=&\dfrac{1}{N}
\llan \Psi^{t} \Phi \rran_{N}.
\end{eqnarray}
The dimensions of $\Phi$ and $\Psi$ are suitably adjusted in the above identities. Similar results can be written for the averaging over $\mathsf{B}$. These are the same results as obtained for CWOE in \cite{vp2010}. However, here we have to take account of $\eta$ which gives further two important identities, viz.,
\begin{eqnarray}\label{Iden5}
\dfrac{1}{T} 
\lan\overline{  \mathsf{A} \Phi \mathsf{B}^{t}\Psi}
\ran_{N}
&=&\llan \Phi \rran_{T} \llan \eta\Psi \rran_{N},
\\
\label{Iden6}
\dfrac{1}{T} 
\lan\overline{  \mathsf{B} \Phi \mathsf{A}^{t}\Psi}
\ran_{M}
&=&\llan \Phi \rran_{T} \llan \eta^{t}\Psi \rran_{M}.
\end{eqnarray}

In this section we omit detail computation of $\overline{G}(z)$, merely stating here the central result of the paper. We provide step by step details of the derivation in Appendix \ref{apB}. A compact result for the self-consistent Pastur equation can be written as
\begin{equation}\label{FinRes}
\overline{G}(z)=
\llan 
\left(z-\zeta\overline{Y}_{1}(z,\overline{G}(z))-\overline{Y}_{2}(z,\overline{G}(z))\right)^{-1}
\rran.
\end{equation}
Here 
\begin{eqnarray}\label{Y1}
\overline{Y}_{1}(z,\overline{G}(z))&=& \dfrac{\left[1+\kappa_{N}\left(z\,\overline{G}(z)-1\right) \right]^{2}}
{1-\dfrac{\kappa_{N}\overline{G}(z)\left[1+\kappa_{N}\left(z\,\overline{G}(z)-1\right) \right]}{1-\kappa_{N}\overline{g}(z,\overline{G}(z))}},
\nonumber
\\
\\
\label{Y2}
\overline{Y}_{2}(z,\overline{G}(z))&=&
\dfrac{\overline{Y}(z,\overline{G}(z))}{1-\kappa_{N}\overline{g}(z,\overline{G}(z))},
\\
\label{Y}
\overline{Y}(z,\overline{G}(z))&=&
\kappa_{M}+\kappa_{N}\left(z\,\overline{G}(z)-1\right)
\nonumber\\
&\times &
\left[1+\kappa_{M}+\kappa_{N}\left(z\,\overline{G}(z)-1\right)
\right],
\end{eqnarray}
and
\begin{equation}\label{gY2}
\overline{g}(z,\overline{G}(z))=\dfrac{[z-\overline{Y}_{2}(z,\overline{G}(z))]\overline{G}(z)-1}{1+\kappa_{N}\left(z\,\overline{G}(z)-1\right)}.
\end{equation}
Eq. (\ref{FinRes}) together with definitions (\ref{Y1}-\ref{gY2}) is the main result of this paper. This result is analogous to the result for CWOE which has been obtained first by Mar\'{c}enko and Pastur \cite{marchenko} and then by others using different techniques \cite{Silverstien,SenM, vp2010}. For the uncorrelated case, i.e., for $\zeta=0$, Eq. (\ref{FinRes}) results in a cubic equation confirming thereby the result obtained in \cite{Burda:2010}.  


\section{Numerical examples and verification of the result (\ref{FinRes})}\label{Numerics}
A numerical technique has been developed in \cite{vp2010} for solving the Pastur equation which describes the density for CWOE. We use the same technique to solve our result (\ref{FinRes}). However, in our case the result is complicated and needs some treatments for meeting requirements of the numerical technique. We first note that $\overline{G}$ depends on $\overline{Y}_{1}$ and $\overline{Y}_{2}$ while both the latter quantities depend on $\overline{g}$ and $\overline{G}$. Since $\overline{g}$ itself depends on $\overline{G}$ and $\overline{Y}_{2}$, at least one quantity $\overline{G}$ or $\overline{g}$, has to be determined explicitly in terms of other. It turns out that, using Eqs. (\ref{gY2}) and (\ref{Y2}), we can eliminate $\overline{Y}_{2}$ from $\overline{g}$. Therefore, for a given $z$, $\overline{g}$, and in turn $\overline{Y}_{1}$ and $\overline{Y}_{2}$, can be estimated using an initial guess of $\overline{G}$. After resolving these we can use the numerical technique \cite{vp2010} to obtain the solution of (\ref{FinRes}).

We demonstrate our result for two different correlation matrices. In a first example we consider $\xi_{\text{AB}}$ to be a rank one matrix, e.g., $[\xi_{\text{AB}}]_{jr}=c$ for every integer $1\le j\le N$ and $1\le r\le M$. In a second example we consider $[\xi_{\text{AB}}]_{jr}=c\,\delta_{jr}+(1-\delta_{jr})c^{|j-r|}$. For simplicity, we consider that the diagonal blocks are defined by equal-cross-correlation matrix model, e.g., $[\xi_{\text{AA}}]_{jk}=\delta_{jk}+(1-\delta_{jk})a$, for $1\le j,k\le N$ and the same for $\xi_{\text{BB}}$ where the correlation coefficient is $b$. In both examples we consider $0<a,b,c<1$. The spectrum of an equal-cross-correlation matrix is easy to calculate. For instance, the spectrum of $\xi_{\text{AA}}$ is described by $1-a$ and $Na+1-a$ where the former has degeneracy $N-1$. The spectrum of $\xi_{\text{BB}}$ is also described in the same way but for $M$ and $b$. The inverse of the square root of these matrices, those we need to define $\eta$, are also not difficult to calculate. For example, we simply have
\begin{eqnarray}
[\xi_\text{AA}^{-1/2}]_{jk}&=&\delta_{jk}[(1-a)^{-1/2}]-\dfrac{1}{N}
\Big[
(1-a)^{-1/2}
\nonumber\\
&-&
(Na+1-a)^{-1/2}
\Big].
\end{eqnarray}

\begin{figure}[!t]
        \centering
               \includegraphics [width=0.4\textwidth]{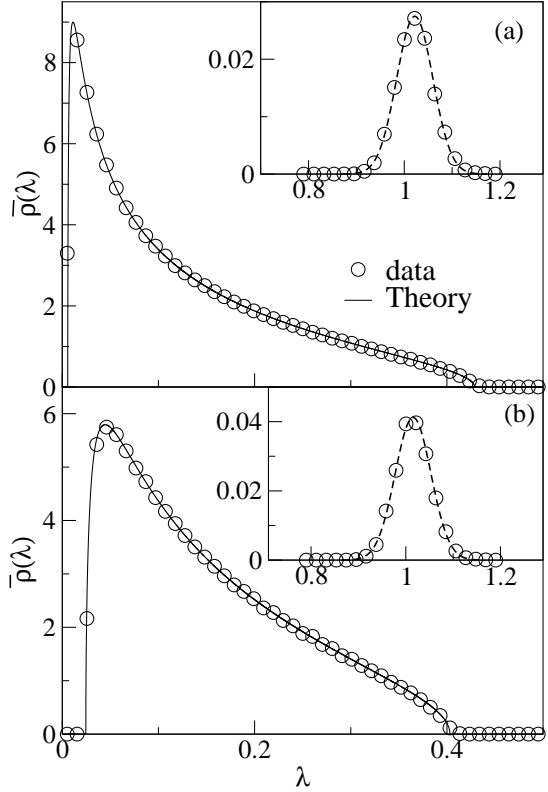}
                \caption{Spectral density, $\overline{\rho}(\lambda)$, where $[\xi_{\text{AB}}]_{jr}=c$ for $1\le j\le N$ and $1\le r\le M$, $c=0.8$, and correlation coefficients of the equal cross-correlation matrices describing the diagonal blocks are $a=b=0.5$. Symbols in the figure represent Monte Carlo simulations and solid lines are the theory obtained from the numerical solution of Eq. (\ref{FinRes}). In Fig. \ref{deneqc}(a) we show results for $N=384$ and in Fig. \ref{deneqc}(b) we show results for $N=256$. The dimension of the full matrix in both the figures is $1024$ and $T=5120$. In the inset we show distribution of the separated eigenvalues where we have considered ensemble of $10000$ matrices. Dashed lines in the inset represent Gaussian distribution where the mean and the variance have been calculated numerically.}
\label{deneqc}
\end{figure}

For the first case the spectrum of $\zeta$ can be determined analytically because of a trivial choice of $\xi_{\text{AB}}$. Yet eigenvalues, $\lambda^{(\xi)}_{j}$, of $\xi$ may not be as trivial to obtain. However, in this case we find $N-1$ eigenvalues $1-a$, $M-1$ eigenvalues $1-b$ and the remaining two are given by
\begin{eqnarray}
\lambda^{(\xi)}_{\pm}=\dfrac{\lambda_{N}^{(\xi_{\text{AA}})}+\lambda_{M}^{(\xi_{\text{BB}})}\pm
\sqrt{
[
\lambda_{N}^{(\xi_{\text{AA}})}-\lambda_{M}^{(\xi_{\text{BB}})}
]^2+4 N M c^{2}
}}{2},
\nonumber\\
\end{eqnarray} 
where $\lambda_{N}^{(\xi_{\text{AA}})}=Na+1-a$ and $\lambda_{M}^{(\xi_{\text{BB}})}=Mb+1-b$. Note that the positive definiteness of $\xi$ is ensured if 
\begin{equation}\label{ineq}
\lambda_{N}^{(\xi_{\text{AA}})}\lambda_{M}^{(\xi_{\text{BB}})}>NM\,c^{2}.
\end{equation}
The matrix $\xi_{\text{AB}}$ is rank one, so is $\eta$:
\begin{eqnarray}
\eta_{jr}&=&\dfrac{c}{\sqrt{\lambda_{N}^{(\xi_{\text{AA}})}\lambda_{M}^{(\xi_{\text{BB}})}}}
\end{eqnarray}
Using this we readily obtain $\zeta_{jk}=Mc^{2}/\lambda_{N}^{(\xi_{\text{AA}})}\lambda_{M}^{\xi_{(\text{BB}})}$ and the only nonzero eigenvalue, $\lambda^{(\zeta)}_{N}=NMc^{2}/\lambda_{N}^{(\xi_{\text{AA}})}\lambda_{M}^{(\xi_{\text{BB}})}$. It trivially follows from the inequality (\ref{ineq}) that $\lambda^{(\zeta)}_{N}<1$, which is actually valid for a more general case as proved in Appendix \ref{apA}.

To check our theoretical result (\ref{FinRes}) with numerics we begin with simulating $\mathcal{C}$ defined in the beginning of the Sec. \ref{secCWE}. Next, we identify the off-diagonal block $\{\text{AB}\}$ in $\mathcal{C}$. Finally we use the transformation $\xi_{\text{AA}}^{-1/2}\mathcal{AB}^{t}\xi_{\text{BB}}^{-1/2}$ to obtain $\mathsf{AB}^{t}$ that we desire to calculate $\mathbf{C}$. In numerical simulations we fix $N+M=1024$, $T=5(N+M)$ and consider two values of $N$, viz., $N=256$ and $384$. To compare the numerics with the theory we consider an ensemble of size $1000$ of matrices $\mathbf{C}$.

\begin{figure}[!t]
        \centering
               \includegraphics [width=0.4\textwidth]{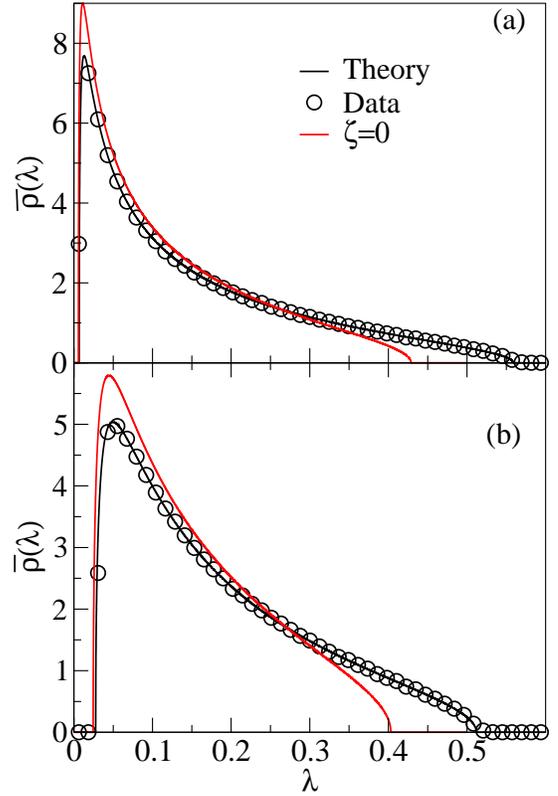}
                \caption{Spectral density, $\overline{\rho}(\lambda)$, for the second example where $[\xi_{\text{AB}}]_{jr}=c\,\delta_{jr}+(1-\delta_{jr})c^{|j-r|}$, for $1\le j\le N$ and $1\le r\le M$, $c=0.05$ and the correlation coefficients which describe the diagonal blocks are $a=b=0.5$. With an outlay similar to Fig. \ref{deneqc} we compare our theory with numerics. Solid red lines in this figure represent the uncorrelated case.}
\label{dencomp}
\end{figure}

In our first example $\zeta$ has only one nonzero eigenvalue. For this spectrum our theory (\ref{FinRes}) yields the density for the bulk of the spectra. It suggests that the bulk should be described by density of the uncorrelated case. We verify this with numerics in Fig. {\ref{deneqc}}, where $a=b=0.9$ and $c=0.8$. However, like the equal-cross-correlation matrix model of CWE \cite{vp2010}, in this case as well, we obtain one eigenvalue separated from the bulk \cite{bulk}. Interestingly, here the bulk remains invariant with correlations as opposed to the CWE case. Moreover, the distribution of the separated eigenvalues is closely described by a Gaussian distribution as shown in insets of Figs. \ref{deneqc}(a) and \ref{deneqc}(b).

Our second example corresponds to a non-trivial spectrum of $\zeta$. We consider the correlation matrix $\xi$ as explained above with parameters  $a=b=0.5$ and $c=0.05$. Note that the off-diagonal blocks have small contributions to the largest eigenvalues of $\xi$ and therefore are difficult to be traced in the analysis of separated eigenvalues of the corresponding CWOE. In Fig. \ref{dencomp} we compare our theory with numerics for $N=384$ in Fig. \ref{dencomp}(a) and for $N=256$ in Fig. \ref{dencomp}(b). As shown in the figure, even small correlations in $\xi_{\text{AB}}$ render notable changes in the density which are described well by our theory.

\section{Conclusion}
In conclusion, we have studied a Wishart model for the nonsymmetric correlation matrices where the two constituting matrices are not statistically independent, incorporating thereby actual correlations in the theory. We have derived a Pastur self-consistent equation which describes the spectral density of this model. Our result is valid for large matrices. We have supplemented some numerical examples to demonstrate the result. 

A couple of interesting analytic problems for this model worth persuing in future, viz., ($i$) to obtain result for spectral density of finite dimensional matrices, and ($ii$) to obtain results for the two-point function and higher order spectral correlations. However, in both the problems calculation of the joint probability density for all the eigenvalues could be a starting point but it seems formidable because of some technicalities. On the other hand, for the unitary invariant ensembles the first problem could be solvable using the techniques of \cite{MousSimon,Baik:2005}. Besides, in the view of success of the supersymmetric method for CWOE \cite{guhr1, guhr2} the first problem seems to be solvable. Moreover, the binary correlation method which has been used to obtain asymptotic result for the two-point function of CWOE \cite{vp2010} could be an effective tool to derive the same for this model. Finally, we believe that given the plenitude of the applications of RMT, these analytic results may not be confined only to time series analysis but in other fields as well \cite{nonhermitian1,nonhermitian2,vmarko}.

\section{Acknowledgments}
The author of this paper is thankful to Thomas H. Seligman for discussions and encouragement. In particular, the author is thankful to F. Leyvraz for useful and illuminating discussions in the course of this work. The author acknowledges referees for invaluable suggestions. 

Financial support from the project 44020 by CONACyT, Mexico, and project PAPIIT UNAM RR 11311, Mexico, in the course of this work is acknowledged. The author is a postdoctoral fellow supported by DGAPA/UNAM. 

\appendix

\section{Upper bound of the singular values of $\eta$}\label{apA}
Since $\xi$ is a positive definite matrix, the matrix $\mathsf{X}$ which results from the decorrelations, defined in Sec. \ref{secCWE}, is also a positive definite matrix. In the following we show that the positive definiteness of $\mathsf{X}$, and therefore of $\xi$, ensures an upper bound of the singular of $\eta$. The matrix $\mathsf{X}$ is given by
\begin{equation}
\mathsf{X}=
\left(
\begin{matrix}
\mathbf{1} & \eta\\
\eta^{t} & \mathbf{1}
\end{matrix}
\right).
\end{equation}
Consider an $(N+M)\times (N+M)$ dimensional orthogonal matrix, $\mathsf{O}$, composed of two orthogonal matrices $\mathbf{O}_{1}$ and $\mathbf{O}_{2}$ of dimensions $N\times N$ and $M\times M$, respectively, defined as
\begin{equation}
\mathsf{O}=
\left(
\begin{matrix}
\mathbf{O}_{1} &0 \\
0& \mathbf{O}_{2} 
\end{matrix}
\right),
\end{equation}
and $\mathbf{O}_{1}\,\eta\,\mathbf{O}_{2}^{t}=\mathsf{S}$ where $\mathsf{S}$ is a rectangular $N\times M$ dimensional diagonal matrix: $\mathsf{S}_{jr}=\delta_{jr}\, s_{j}$ and the $s_{j}$'s are the singular values of $\eta$. Then 
\begin{equation}
\mathsf{OXO}^{t}=
\left(
\begin{matrix}
\mathbf{1}  & \mathsf{S}\\
\mathsf{S^{t}} & \mathbf{1}
\end{matrix}
\right).
\end{equation} 
Since $\mathsf{X}$ is a positive definite matrix, therefore $\mathsf{OXO}^{t}$ is also a positive definite matrix. We use the Sylvester's criterion \cite{Sylvester} which states that a real symmetric matrix is positive definite iff all the leading principal minors of the matrix are positive. This criterion, for $\mathsf{OXO}^{t}$, leads to $n$ number of inequalities where $n=\text{min}\{N,M\}$. For instance, for $N<M$ we have $N$ inequalities: $\prod_{k}^{N-j}(1-s_{k}^{2})>0$, for $j=0,...,N-1$. These inequalities hold together if $s_{j}<1$, for all the $j$'s, giving thereby an upper bound $1$ due to the positive definiteness of $\mathsf{OXO}^{t}$ and therefore due to the positive definiteness of $\mathsf{X}$.


\section{Derivation of the result (\ref{FinRes})}\label{apB}
We prefer to calculate a more general quantity $\overline{G}_{L}$, defined as
\begin{equation}
\overline{G}_{L}(z)=\llan\overline{ \, L \dfrac{1}{z  \textbf{1}_{N\times N} -\mathbf{C}}}\,\rran.
\end{equation}
Here $L$ is an arbitrary but nonrandom $N \times N$ matrix. What follows from the identities (\ref{Iden1}-\ref{Iden6}) is that the binary associations of $\mathsf{A}$ only with $\mathsf{A}^{t}$ or with $\mathsf{B}^{t}$ give leading order terms, otherwise $\mathcal{O}(N^{-1})$ or lower order terms. Therefore, in the expansion (\ref{ResMp}), we calculate only the binary associations described below.
\begin{eqnarray}\label{demobin}
&&\overline{G}_{L}(z)=\dfrac{\llan L\rran}{z}+\sum_{p=1}^{\infty}
\dfrac{z^{-p-1}}{T^{2}}
\Big\{
\lan\overline{  L\,\underbracket[1pt]{\mathsf{AB}^{t}}\mathsf{BA}^{t} \mathbf{C}^{p-1}} \ran
\nonumber\\
&+&
\lan\overline{  L\underbracket[1pt]{\mathsf{AB}^{t}\mathsf{BA}^{t}} \mathbf{C}^{p-1} }\ran
\Big\}
+\sum_{p=2}^{\infty}\sum_{n=0}^{p-2}
\dfrac{z^{-p-1}}{T^{4}}
\nonumber\\
&\times&
\Big\{
\lan\overline{ 
L\, \underbracket[1pt]{\mathsf{AB}^{t}\mathsf{BA}^{t}\mathbf{C}^{n}\mathsf{AB}^{t}}\mathsf{BA}^{t}
\mathbf{C}^{p-n-2}}
\ran
\nonumber\\
&+&
\lan\overline{ 
L\, \underbracket[1pt]{\mathsf{AB}^{t}\mathsf{BA}^{t}\mathbf{C}^{n}\mathsf{AB}^{t}\mathsf{BA}^{t}} 
\mathbf{C}^{p-n-2}}
\ran
\Big\}.
\end{eqnarray}  
Here we avoid terms due to binary associations of $\mathsf{A}$ with $\mathsf{A}$ or with $\mathsf{B}$ since the former are $\mathcal{O}(N^{-1})$, because of the identity (\ref{Iden2}), and the latter vanish on ensemble averaging. The binary associations we consider in (\ref{demobin}) then yield a leading order equality. Using the identities (\ref{Iden1},\ref{Iden5}) we get
\begin{eqnarray}\label{GgD}
&&\overline{G}_{L}(z)=\dfrac{\llan L\rran}{z}+
\dfrac{\overline{g}_{L}(z)}{z}\left\{ 1+\kappa_{N}\left(z\,\overline{G}(z)-1\right)\right\}
\nonumber\\
&+&
\dfrac{\kappa_{M}}{z}\,\overline{G}_{L}(z)+
\kappa_{M}\,\overline{G}_{L}(z)\,\sum_{n=1}^{\infty}
\dfrac{\lan\overline{  \mathbf{D}^{n}\mathsf{BB}^{t}}\ran}
{T\,z^{n+1}},
\end{eqnarray}
where we have used definition (\ref{D}) to write the last term in the right hand side and
\begin{eqnarray}\label{defgL}
\overline{g}_{L}(z)&=&\sum_{p=0}^{\infty}\dfrac{\lan\overline{  L\, \eta\mathsf{BA}^{t}\mathbf{C}^{p}}\ran}{T\,z^{p+1}}.
\end{eqnarray}
It should be mentioned that the angular brackets we are using for the spectral averaging are in accordance with the dimensionality of the matrices under trace operation. For instance the spectral average in the term, involving $\mathbf{D}$, of Eq. (\ref{GgD}) is calculated over an $M\times M$ matrix. Binary associations across the traces, in intermediate steps from (\ref{demobin}) to (\ref{GgD}), are also ignored as they produce lower order terms; see identities (\ref{Iden3},\ref{Iden4}).   
 
To calculate the summation in Eq. (\ref{GgD}) we consider the binary associations similar to those in Eq. (\ref{demobin}), but for $\mathsf{B}$. We obtain
\begin{eqnarray}
z\kappa_{M}\sum_{n=1}^{\infty}
\dfrac{\lan\overline{  \mathbf{D}^{n}\mathsf{BB}^{t}}\ran}{T\,z^{n+1}}
&=&
\Big[
\kappa_{N}\left(z\,\overline{G}(z)-1\right)
\nonumber\\
&\times&
\left[1+\kappa_{N}\left(z\,\overline{G}(z)-1\right)+\kappa_{M}
\right]
\nonumber\\
&+&\kappa_{N}\kappa_{M}\overline{g}(z)
\Big]\times \left[1-\kappa_{N}\overline{g}(z)\right]^{-1}.
\nonumber\\
\end{eqnarray}  
In the derivation of the above equation we have used the resolvent $\mathcal{G}(z)$ defined for $\mathbf{D}$ as $\mathcal{G}(z)=\llan(z\textbf{1}_{M\times M}-\mathbf{D})^{-1}\rran$. Since $\mathbf{D}$ and $\mathbf{C}$ have the same nonzero spectrum, $\overline{\mathcal{G}}(z)$ can be easily given in terms of $\overline{G}(z)$, as
\begin{equation}
\overline{\mathcal{G}}(z)-z^{-1}=\dfrac{\kappa_{N}}{\kappa_{M}}\left(\overline{G}(z)-z^{-1}\right).
\end{equation}
Finally, we have used the fact that for a square matrix the trace remains the same for its transpose. 

However, we still remain with $\overline{g}_{L}(z)$. Considering again the binary associations of $\mathsf{B}$, in (\ref{defgL}), we find 
\begin{eqnarray}
\overline{g}_{L}(z)&=&
\overline{G}_{L\zeta}(z)\,
\left\{1+\kappa_{N}\left(z\,\overline{G}(z)-1\right)\right\}
\nonumber\\
&+&
\kappa_{N}\overline{g}_{L}(z)
\sum_{n=0}^{\infty}\dfrac{\lan\overline{  \mathsf{AA}^{t}\mathbf{C}^{n}}\ran}{T\,z^{n+1}}.
\end{eqnarray}
Using the binary associations of $\mathsf{A}$, the left over summation is computed to be
\begin{eqnarray}
\sum_{n=0}^{\infty}\dfrac{\lan\overline{  \mathsf{AA}^{t}\mathbf{C}^{n}}\ran}{T\,z^{n+1}}
&=&
\dfrac{\overline{G}(z)\left[1+\kappa_{N}\left(z\,\overline{G}(z)-1\right) \right]}
{1-\kappa_{N}\overline{g}(z)}.
\end{eqnarray}
It readily gives $\overline{g}_{L}$ in terms of $\overline{g}(z)$ and $\overline{G}(z)$ as
\begin{equation}\label{gl}
\overline{g}_{L}(z,\overline{G}(z))=
\dfrac{\overline{G}_{L\zeta}(z)\left[1+\kappa_{N}\left(z\,\overline{G}(z)-1\right) \right]}
{1-\dfrac{\kappa_{N}\overline{G}(z)\left[1+\kappa_{N}\left(z\,\overline{G}(z)-1\right) \right]}{1-\kappa_{N}\overline{g}(z,\overline{G}(z))}}.
\end{equation}

We can now rewrite the equation (\ref{GgD}) in a closed form. For instance, we write
\begin{equation}\label{GY1Y2}
z \overline{G}_{L}(z)=\llan L\rran+
\overline{G}_{L\zeta}(z)\,\overline{Y}_{1}(z,\overline{G}(z))+\overline{G}_{L}(z)\,\overline{Y}_{2}(z,\overline{G}(z)),
\\
\end{equation}
where $\overline{Y}_{1}(z,\overline{G}(z))$ and $\overline{Y}_{2}(z,\overline{G}(z))$ are defined respectively in Eq. (\ref{Y1}) and Eq. (\ref{Y2}). Substituting $L\rightarrow L[z-\zeta\,\overline{Y}_{1}(z,\overline{G}(z)) -\overline{Y}_{2}(z,\overline{G}(z))]^{-1}$, in Eq. (\ref{GY1Y2}), we obtain 
\begin{equation}\label{ResGL}
\overline{G}_{L}(z)=
\llan L\, \dfrac{1}{
z-\zeta \overline{Y}_{1}(z,\overline{G}(z)) -\overline{Y}_{2}(z,\overline{G}(z))}
\rran.
\end{equation}
For $L=\textbf{1}_{N\times N}$ the above equation gives the central result (\ref{FinRes}) of the paper.

Using Eq. (\ref{gl}), for $L=\textbf{1}_{N\times N}$, and definition (\ref{Y1}), we write $\overline{g}(z,\overline{G}(z))$ as
\begin{equation}
\overline{g}(z,\overline{G}(z))=\dfrac{\overline{G}_{\zeta\,\overline{Y}_{1}(z,\overline{G}(z))}(z)}{1+\kappa_{N}\left(z\,\overline{G}(z)-1\right)}.
\end{equation}
For $L=\zeta\, \overline{Y}_{1}(z,\overline{G}(z))$, it is straightforward to deduce the definition (\ref{gY2}) from Eq. (\ref{ResGL}).



\begin{thebibliography}{99}
\bibitem{bulk} It is understood that in both cases the bulk density is normalized to $1-N^{-1}$.
\end{thebibliography}
\end{document}